\documentclass[prl,
twocolumn,
superscriptaddress,showpacs,amsmath,amssymb]{revtex4}
\usepackage{amsfonts}
\usepackage{bm}
\usepackage{verbatim}
\usepackage{color}
\usepackage{graphicx}
\usepackage[applemac]{inputenc}

\newcommand{\abs}[1]{\left \vert #1\right \vert}
\newcommand{\im}{i}
\newcommand{\unitvec}[1]{\hat{\mathbf{#1}}}
\newcommand{\vek}[1]{\mathbf{#1}}
\newcommand{\doo}{\partial}
\newcommand{\hc}{\textrm{ h.c.}}

\begin{document}

\title{Emergent spacetime and gravitational Nieh-Yan anomaly in chiral $p+ip$ Weyl superfluids and superconductors}

\author{Jaakko Nissinen}

\affiliation{Low Temperature Laboratory, Department of Applied Physics, Aalto University, P.O. Box 15100, FI-00076
Aalto, Finland}

\email{jaakko.nissinen@aalto.fi}

\date{\today}

\begin{abstract}
Momentum transport is anomalous in chiral $p+ip$ superfluids and superconductors in the presence of textures and superflow. Using the gradient expansion of the semi-classical approximation, we show how gauge and Galilean symmetries induce an emergent curved spacetime with torsion and curvature for the quasirelativistic low-energy Majorana-Weyl quasiparticles. We explicitly show the emergence of the spin-connection and curvature, in addition to torsion, using the superfluid hydrodynamics. The background constitutes an emergent quasirelativistic Riemann-Cartan spacetime for the Weyl quasiparticles {which} satisfy the conservation laws associated with local Lorentz symmetry restricted to the plane of uniaxial anisotropy of the superfluid (or -conductor). Moreover, we show that the anomalous Galilean momentum conservation is a consequence of the gravitational Nieh-Yan (NY) chiral anomaly the Weyl fermions experience on the background geometry. Notably, the NY anomaly coefficient features a non-universal ultraviolet cut-off scale  $\Lambda$, with canonical dimensions of momentum. Comparison of the anomaly equation and the hydrodynamic equations suggests that the value of the cut-off parameter $\Lambda$ is determined by the normal state Fermi liquid and non-relativistic uniaxial symmetry of the $p$-wave superfluid or superconductor.
\end{abstract}

\maketitle

\emph{Introduction. ---} Topological phases can be classified in terms of quantum anomalies that are robust to interactions and other perturbations \cite{RyuEtAl12, Wen2013, Kapustin2014}. Protected emergent quasi-relativistic {Fermi} excitations coupled to gauge fields and geometry arise as dictated by topology and anomaly inflow \cite{CallanHarvey85,Volovik2003, Horava05}. In particular gapless fermions with Weyl spectrum and chiral anomalies are a recent prominent example \cite{AdlerBellJackiw, NielsenNinomiya83, Volovik85,Volovik1986a, BalatskiiEtAl86, Volovik87, WanEtAl11, ZyuzinBurkov12, SonYamamoto12, Zaheed2012, SonSpivak13}. On the other hand, topological phases and their coupling to geometry (and gravity) is currently a rapidly advancing subject. The well-established results concern the Hall viscosity \cite{AvronSeilerZograf95} and chiral central charge  \cite{Volovik1986b, Volovik90, ReadGreen00} related to gravitational anomalies \cite{EguchiFreund76, AlvarezGaumeWitten84, AlvarezGaumeGinsparg85} and thermal transport in quantum Hall systems, topological superfluids (SFs) and superconductors (SCs), as well as semimetals \cite{Read09, Landsteiner11, Stone12,  HoyosSon12, BasarKharzeevZahed2013, BradlynEtAl12, HughesEtAl13, CanLaskinWiegmann14, AbanovGromov14, ParrikarEtAl14, SunWan2014, Zubkov2015, Valle2015, BradlynRead15, GromovEtAl15, GromovAbanov15, Lucas2016, PalumboPachos16, GoothEtAl17, Wiegmann18, FujiiNishida18, KvorningEtAl18, GolanStern18, KhaidukovZubkov2018, MaranerPachosPalumbo18, GromovEtAl19, GolanHoyosMoroz19, CopettiLandsteiner19, FerreirosEtAl19, HuangEtAl19, NissinenVolovikPRR19}, the case of SFs and SCs being especially important due to the lack of conserved charge. Any purported topological response of geometrical origin is necessary more subtle than that based on gauge fields with conserved charges due to the inherent dichotomy between topology and geometry.

Topologically protected Weyl quasiparticles arise also in three-dimensional chiral $p$-wave SFs and SCs {from the gap nodes} \cite{Volovik2003, Volovik92} {and lead to non-zero normal density close to the Fermi surface zeros, even at zero temperature \cite{VolovikMineev81, CombescotDombre86}}. As any chiral fermion in three dimensions, they suffer from the chiral anomaly in the presence of non-trivial background fields, now as an anomaly where momentum is transferred from the order parameter fluctuations to the quasiparticles \cite{MerminMuzikar80, VolovikMineev81, CombescotDombre83, CombescotDombre84, Volovik84, CombescotDombre86, Volovik92},
\begin{align}
\doo_t \mathbf{P}_{\rm vac} + \nabla \cdot \boldsymbol{\Pi}_{\rm vac} = -\doo_t \mathbf{P}_{\rm qp}-\nabla \cdot \boldsymbol{\Pi}_{\rm qp} \neq 0, \label{eq:momentum_anomaly}
\end{align}
where $\mathbf{P}_{\rm vac, qp}$ and $\boldsymbol{\Pi}_{\rm vac, qp}$ refer, respectively, to the momentum and stress-tensor of the order parameter vacuum and quasiparticles. {Due to textures and superflow, the quasiparticles and -holes flow through the gap nodes transferring net momentum}. The total momentum is conserved. The chiral anomaly \eqref{eq:momentum_anomaly} on vortices \cite{Volovik95} has been measured in the early landmark experiment {in $^3$He-A} \cite{BevanEtAl97}. {See the Supplementary Materials (SM) for a detailed review of Eq. \eqref{eq:momentum_anomaly} in terms of SF hydrodynamics}. 

{The anomaly \eqref{eq:momentum_anomaly} relates to the famous angular momentum ``paradox" of the chiral superfluid/superconductor \cite{McClureTakagi79, MerminMuzikar80, VolovikMineev81, CombescotDombre83, CombescotDombre84, IshiharaEtAl19}: even though each Cooper pair carries net angular momentum $\hbar$, the pairs overlap substantially. In a fixed volume, the overlap is of the order $\sim\! a^2/\xi^2$ with mean pair-separation $a$ and coherence length $\xi$. This makes the local angular momentum contribution very small, of the order $\sim\! (\Delta_0/E_F)^2$ instead of (half) the total fermion density, where $\Delta_0$ is the gap and $E_F$ the Fermi energy. At equilibrium, the local variation of angular momentum is still well-defined. Similarly, the total linear momentum density is well-defined but it is no longer conserved separately between the condensate and normal component, producing the anomaly \cite{VolovikMineev81, CombescotDombre86}. This is in contrast to a system of non-overlapping Bose Cooper pair ``molecules" where the anomaly due to Weyl nodes naturally also vanishes \cite{VolovikMineev81, Volovik92}.}

Here we show that the momentum anomaly \eqref{eq:momentum_anomaly} is a manifestation of the so-called chiral gravitational Nieh-Yan (NY) anomaly \cite{NiehYan82, NiehYan82b,Yajima96, ObukhovEtAl97, ChandiaZanelli97, Soo99, PeetersWaldron99} on emergent spacetime with torsion (and curvature) coupling to the low-energy Weyl fermions. The subtle role of this {gravitational (geometric)} anomaly term {due to torsion} has been debated in the literature since its discovery \cite{Comments}. We show how it arises through the Galilean symmetries and hydrodynamics of the non-relativistic system with an explicit ultraviolet completion. Taken at face value, the NY anomaly has therefore been experimentally verified in the late 90's, albeit in the context of emergent condensed matter fermions and spacetime induced by the SF order parameter. Emergent spacetime in two-dimensional topological SCs was recently carefully discussed in Ref. \onlinecite{GolanStern18}. Related recent work discusses emergent tetrads, gauge fields and anomaly terms in Weyl SFs \cite{KobayashiEtAl18, IshiharaEtAl19} and semimetals \cite{GrushinEtAl16, PikulinEtAl16, WeststromOjanen17, FerreirosEtAl19, LiangOjanen19, IlanEtAl19}, without the framework of emergent conservation laws and geometry.
 
\emph{Model. ---} We consider an equal spin pairing $p$-wave SF at zero-temperature on flat Euclidean space in the mean-field (MF) approximation \cite{VollhardtWoelfle, Volovik2003} {(see the SM for more details)}. We comment below how to extend our results to SCs. The action for spinless Grassmann fermion $\{ \Psi(x)$, $\Psi^{\dagger}(x')\} = 0$ is ($\hbar = 1$, summation over repeated indices)
\begin{align}
S[\Psi,\Psi^{\dagger}, \Delta, \Delta^\dagger] &= \int d^4 x~ \Psi^\dagger \im\doo_t \Psi - \mathcal{H}_{\rm MF}, \nonumber\\ 
\mathcal{H}_{\rm MF} = \epsilon(\Psi,\doo_i\Psi)&+\frac{1}{2\im}( \Delta^i \Psi^\dagger \doo_i \Psi^\dagger + \Delta^{* i} \Psi \doo_i \Psi ). \label{eq:action}
\end{align}
The normal state energy is $\epsilon(\Psi, \doo_i \Psi) = \frac{\doo_i\Psi^\dagger \doo_i \Psi}{2m} - \mu_F \Psi^\dagger \Psi$, where $m$ is the constituent mass, $\mu_F=p_F^2/2m$ is the normal state Fermi level. The {spinless} MF gap amplitude $\Delta^i(x) \propto \frac{1}{2\im}\langle \Psi \doo_i \Psi\rangle$ is {$\Delta^i =  \frac{\Delta_0}{p_F}(\hat{\mathbf{m}}-i\hat{\mathbf{n}})^i \equiv c_{\perp}(\hat{e}_1-i\hat{e}_2)^i$}, the unit vector $\hat{e}_3 = \unitvec{m}\times \unitvec{n} = \unitvec{l}$ being the axis of orbital angular momentum of the Cooper pairs. The dynamics of the SF free-energy is ignored and {$\Delta(x), \Delta^{\dagger}(x)$} represent given background fields. We ignore Fermi liquid corrections \cite{VollhardtWoelfle} for simplicity as we expect that these will not affect the arguments which are based on the symmetries, hydrodynamics and anomalies of the system. For SCs, we {perform minimal substitution in $\epsilon(\Psi, \doo_i\Psi)$}.

\emph{Symmetries and Galilean transformations. --- } The unbroken continuous symmetries of the normal Fermi liquid state are U(1)$_N$ $\times$ $\mathbb{R}^3$ $\times$ SO(3)$_{L}$, where the translations and rotations $\mathbb{R}^3$ $\times$ SO(3)$_{L} $ form a subgroup of the Galilean group. The gauge and rotational symmetries are broken to the combined gauge symmetry U(1)$_{N/2+L_3}$ \cite{LiuCross79}. In addition, time-reversal symmetry is broken allowing for the emergence of Weyl quasiparticles. 

In the SF, the global U(1)$_N$ gauge symmetry {leads to} the conservation law
$\doo_{\mu}J^{\mu} =  -\Delta^i \Psi^{\dagger} \doo_i\Psi^{\dagger} + \Delta^{*i} \Psi\doo_i\Psi$, where {$J^{\mu} = (\rho, J^i)$} is the normal state fermion current. 
A Galilean {transformation from the SF comoving frame (c.f.m.)} is given as $\mathbf{x}' = \mathbf{x}+\mathbf{v}_s t$ and $t' = t$. In terms of SF velocity $\mathbf{v}_s$ and chemical potential $\mu_m$, we transform \cite{VollhardtWoelfle, SonWingate06}
\begin{align}
\Psi(x) &\to \Psi'(x') = e^{-\im [m\mathbf{v}_s\cdot \mathbf{x}+\frac{1}{2}(m \mathbf{v}_s^2 - 2\mu_m)t]}\Psi(\mathbf{x}+\mathbf{v}_s t,t), \nonumber\\
\Delta(x) &\to e^{-\im[2 m \mathbf{v}_s\cdot \mathbf{x}+(m\mathbf{v}_s^2-2 \mu_m)t]}\Delta(\mathbf{x}+\mathbf{v}_s t,t) . \label{eq:Galilean_trans}
\end{align}
The gauge {and} Galilean transformations are not independent {for} coordinate dependent transformation parameters. {For} infinitesimal constant velocity, the action changes in the c.m.f. with $\Psi'(x')$ as $\delta S_{\rm c.m.f.}[\mathbf{v}_s,\mu_m]
= \int d^4 x~ m\mathbf{v}_{s} \cdot J^{i} - \mu_m \rho + O(\mathbf{v}_s^2)$. Equivalently  $\doo_{\mu}\Psi \to (\doo_{\mu}-\im m v_{\mu})\Psi$, where $v_{\mu} = (-\mu_m/m, \mathbf{v}_s)$. Rotations {along} $\hat{e}_3$ {act as $\Delta^{i} \to e^{\im\varphi}\Delta^{i}$ and lead to the combined gauge symmetry \cite{LiuCross79}.} {The SF velocity is determined as}
\begin{align}
2mv_{s}^i = -\hat{e}_1\cdot \doo_{i} \hat{e}_2 = \doo_i\varphi, 2 \mu_m = \hat{e}_1\cdot \doo_t \hat{e}_2 = -\doo_t \varphi,
\end{align}
where $-\varphi$ is the rotation angle. In addition the SF velocity satisfies the Mermin-Ho relations \cite{MerminHo76, Volovik92}
\begin{align}
\nabla \times \mathbf{v}_s = -\frac{\kappa}{4 \pi}\epsilon^{ijk} \hat{\mathbf{l}}_i  \nabla \hat{\mathbf{l}}_j \times \nabla \hat{\mathbf{l}}_k, \nonumber \\
\doo_t \mathbf{v}_s + \nabla\mu_m =  -\frac{\kappa}{2\pi} \epsilon^{ijk}\unitvec{l}_i \partial_t \unitvec{l}_j \nabla \unitvec{l}_k . \label{eq:Mermin-Ho}
\end{align}
where $\kappa = h/2m$ is the circulation quantum. The translation symmetry corresponds to the {energy-momentum tensor} conservation law,  {$S =\int d^4 x \mathcal{L}$},
\begin{align}
\doo_{\mu}\Pi^{\mu}_{\nu} = \doo_{\nu}\mathcal{L}\vert_{\rm explicit}, \label{eq:momentum_conservation}
\end{align}
where $\Pi^{\mu}_{\nu} = -\frac{\doo \mathcal{L}}{\doo(\doo_{\mu}\Psi)}\doo_{\nu} \Psi + \hc + \mathcal{L}\delta^{\mu}_{\nu}$. 
{In particular, this conservation law is broken by the anomaly \eqref{eq:momentum_anomaly} for $\nu=i$ {along $\unitvec{l}$}}. Moreover, $\Pi^i_j$ is not symmetric in the plane determined by $\Delta^i$, as is well known in $^3$He-A \cite{Volovik92, VollhardtWoelfle, GolanStern18}. {See the SM for more details on the symmetries of the model}.

\emph{Linearized comoving quasiparticle action. ---}
We define the Bogoliubov {transformation} $\Psi(x) = \sum_s u_s(x) a_s +v_{s}^*(x) a_{s}^{\dagger}$, where $s$ is a generalized index in quasiparticle (particle-hole) space with $\{a^\dagger_s, a_s'\} = \delta_{ss'}$. The Bogoliubov-de Gennes (BdG) quasiparticles form a spinless two-component Grassman Nambu spinor $\Phi(x) \sim\sum_s \left(\begin{matrix}u_s(x) & v_s(x)\end{matrix} \right)^Ta_s$, with the Lagrangian
\begin{align}
\mathcal{L}_{\rm BdG}
= \Phi^{\dagger}(x) \left( \begin{matrix} \im\doo_t -\epsilon(-\im \doo_i) & \frac{1}{2}\{\Delta^i, \im \doo_i\} \\ \frac{1}{2}\{\Delta^{*i}, \im \doo_i\} & \im\doo_t+\epsilon(\im\doo_i) \end{matrix} \right) \Phi(x) . 
\end{align}
If $(u_s \ v_s)$ is a solution to the equation of motion with energy $\varepsilon_s$, $(v^*_{-s} \ u_{-s}^*)$ satisfies particle-hole conjugate solution $\varepsilon_{-s} = -\varepsilon_{s}$. This implies the Majorana relation $\tau^1 \Phi^{\dagger} \sim \Phi$. {For a homogenous state, the dispersion vanishes on the Fermi surface at $\mathbf{k}=\pm p_F\unitvec{l}$}. The same BdG action applies for SCs with the replacements $\doo_{t} \to \doo_{t} -\im A_{0}\tau^3$, $\tau_3\epsilon(\mp\im\doo_i) \to \tau^3 \epsilon(\mp(\im\doo_i+A_{i}))$, where the signs are according to eigenstates of $\tau^3$ and $v_{\mu} \to v_{\mu} - A_{\mu}$, where $A_{\mu}$ is the electromagnetic gauge potential.

Now we consider the low-energy fermions in the presence of slowly varying order parameter texture and Galilean transformation parameters $\mathbf{v}_s(x), \mu_m(x)$ in the semiclassical approximation, see e.g. \cite{Kopnin}. We assume that the BdG fermions with gap $\gtrsim mc^2_{\perp}$ have been integrated out and restrict ourselves to the linear expansion close to the Weyl nodes $\pm p_F\unitvec{l}$. The dispersion is
\begin{align}
\Phi^{\dagger}(x) \epsilon(-\im\doo) \Phi(x) \approx \sum_{\pm} \tilde{\chi}_{\pm}^{\dagger}(x)\{\frac{v_F}{2}\unitvec{l}^i(x), -\im \doo_i\} \tilde{\chi}_{\pm}(x) \label{eq:node_trans} 
\end{align}
and we neglect terms of order $O(\doo^2 \tilde{\chi})$, where $\Phi(x) = \sum_{\pm}e^{\pm \im p_F \int^{\mathbf{x}} \unitvec{l}(\mathbf{x}') \cdot d\mathbf{x'}}\tilde{\chi}_{\pm}(x)$ and $\tilde{\chi}_{\pm}(x)$ is slowly varying compared to $p_F$. The order parameter vectors form the spatial part of an inverse tetrad $e^{\ i}_a = \{(\frac{c_{\perp}}{c_{\parallel}})\hat{\mathbf{m}}, (\frac{c_{\perp}}{c_{\parallel}})\hat{\mathbf{n}}, \hat{\mathbf{l}}\}$ with uniaxial symmetry. {Note that after linearization of $\epsilon(-\im\doo)$, we perform in \eqref{eq:node_trans} the transformation $\Phi \to \tilde{\chi}$, i.e. $\{ e^i_3, (-\im\doo_i-\vek{e}^3 p_F)_i \} \to \{e^{i}_3 ,\doo_i\}$ corresponding to a chiral rotation in momentum space {due to the node}. This is precisely the anomalous chiral symmetry in the system and produces the quasirelativistic anomaly \eqref{eq:NYanomaly} below, proportional to $\Lambda \propto p_F$ in the path-integral representation \cite{Fujikawa79, ChandiaZanelli97, Soo99}.}

Further, in the presence of SF velocity $v_{\mu} =\doo_{\mu}\varphi  = 2m (\mu_m/m,-\mathbf{v}_s)$, {the fermions} transform in the comoving frame as $\tilde{\chi} \to e^{-\im \varphi \tau^3/2}\tilde{\chi}$, $e^i_1-\im e_2^i \to e^{-\im \varphi}(e_{1}-\im e_2)^i$ where $\varphi(x)$ is slowly varying compared to $p_F$. The derivative operator transforms to {$\tau^a \im e^{\mu}_a \doo_{\mu} \to \tau^a \im e_a^{\mu} (\doo_{\mu} - \frac{\im}{2}\doo_{\mu}\varphi\tau^3)$}. These coincide with local spin-1/2 Lorentz transformations in the 12-plane, and we can attempt to associate the linearized action to a non-trivial spacetime. Denoting the Pauli matrices $\tau^a, \overline{\tau}^a = (1,\pm\tau^i)$ in Nambu space, velocities $c_{\parallel} = v_F = p_F/m$ and $c_{\perp} = \frac{\Delta_0}{p_F}$ and $\doo_{\mu} = (\doo_t, c_{\parallel}\doo_i)$, the linearized {action} close to the node $+p_F\unitvec{l}$, written in explicitly hermitean form, becomes
\begin{widetext}
\begin{align}
S[\tilde{\chi}^\dagger, \tilde{\chi},\Delta, \mu_m,\mathbf{v}_s ] &= \int d^4 x \frac{1}{2} (\tilde{\chi}^\dagger \im \doo_t \tilde{\chi} -\im \doo_t \tilde{\chi}^\dagger\tilde{\chi}) + (\mu_F - \mu_m) \tilde{\chi}^\dagger \tau^3\tilde{\chi} - \frac{p_F\unitvec{l}^i}{2} \tilde{\chi}^\dagger v_F\unitvec{l}^i \tau^3 \tilde{\chi} \label{eq:linear_action}\\ + \frac{v_F\unitvec{l}^i}{2}(\tilde{\chi}^\dagger \tau^3 \im \doo_i\tilde{\chi} & - \im \doo_i\tilde{\chi}^\dagger \tau^3\tilde{\chi} )  - \frac{1}{2} (\tilde{\chi}^\dagger (e^i_1\tau^1 \im \doo_i + e^i_2\tau^2\im \doo_{i})\tilde{\chi} +\hc  ) + p_F \unitvec{l}\cdot \mathbf{v}_s  \tilde{\chi}^\dagger\tilde{\chi} -  \frac{v_s^i}{2}
(\tilde{\chi}^\dagger \im \doo_i \tilde{\chi} -\im \doo_i\tilde{\chi}^\dagger\tilde{\chi}). \nonumber
\end{align}
\end{widetext}
\emph{Emergent Riemann-Cartan spacetime. ---} The action \eqref{eq:linear_action} is (after a rotation of $\tau^a$ in the 12-plane) equivalent to a relativistic chiral (right-handed) Majorana-Weyl fermion on Riemann-Cartan spacetime \cite{NiehYan82b, Shapiro02, Hammond}, see the SM for a review of Rieman-Cartan spacetimes  and our conventions on relativistic fermions,
\begin{align}
S_{\rm Weyl}[\chi, \chi^\dagger, e, \omega] = \frac{1}{2} \int e d^4x~  \chi^\dagger \tau^a e^{\mu}_a \im \overset{\rightarrow}{D}_{\mu} \chi + \textrm{h.c.} \label{eq:Weyl_action}
\end{align}
with the identifications $\chi = e^{-1/2}\tilde{\chi}$, $e = \det e^a_{\mu} =\tfrac{c_{\parallel}^2}{c_{\perp}^2}$,
\begin{align}
e_0^{\mu} &= (1,- \mathbf{v}_s), \quad e^{\mu}_1 = \tfrac{c_{\perp}}{c_{\parallel}}(0, \unitvec{m}), \nonumber\\
e^{\mu}_2 &=  \tfrac{c_{\perp}}{c_{\parallel}}(0, \unitvec{n}), \quad e^{\mu}_3 = (0, \unitvec{l}) \label{eq:tetrad}
\end{align}
and $\omega_{\mu}^{12}= 2mv_{s\mu} = 2m (-\mu_m/m, \mathbf{v}^i_s) = \doo_{\mu}\varphi$. The covariant derivate is $\overset{\rightarrow}{D}_{\mu} = \doo_{\mu}-\frac{\im}{4} \omega_{\mu}^{ab}\sigma_{ab}$, $\sigma^{ab} = \frac{\im}{2}[\overline{\tau}^a, \tau^b]$. Note that the Galilean invariance {leads to both the shift $e_0^{i} = -v^i_s$, although it is $O(\doo^2)$ in the action, and the spin-connection} term $e^i_3\omega_{i}^{12}\tilde{\chi}^\dagger \tilde{\chi} \propto p_F\unitvec{l}\cdot \mathbf{v}_s$ in Eq. \eqref{eq:linear_action}. We {also emphasize that we explicitly} retained the {momemtum density} $p_F^2 \unitvec{l}^2/2m \tilde{\chi}\tau^3\tilde{\chi}$ in the action, which cancels with the Fermi level $\mu_F\tilde{\chi}\tau^3\tilde{\chi}$, since {this term} represents non-zero physical Galilean momentum present at the Weyl node, {see Eq. \eqref{eq:lin_energy_momentum} below}.

The Eqs. \eqref{eq:Weyl_action}, \eqref{eq:tetrad} and the spin-connection $\omega^{12}_{\mu}$ define a Weyl fermion on an emergent Riemann-Cartan spacetime \cite{NiehYan82b, Shapiro02}, with the uniaxial metric $g^{\mu\nu} = e_a^{\mu} e^{\nu}_b \eta^{ab}$, $e^{a}_{\mu}e^{\nu}_a = \delta^{\mu}_{\nu}$ and $\nabla_\mu e^a_{\nu} \equiv \doo_{\mu}e^a_{\mu} + \omega^a_{\mu b}e^a_{\nu} - \Gamma^{\lambda}_{\mu\nu}e^a_{\lambda} = 0$. The {non-zero} torsion and curvature tensors are given by $T^a_{\mu\nu} = \doo_{\mu}e^a_{\nu} - \doo_{\nu}e^a_{\mu} + \omega_{\mu b}^{a}e^b_{\nu} -\omega_{\nu b}^{a}e^b_{\mu}$ and {$R^{12}_{\mu\nu} = \doo_{\mu}\omega_{\nu}^{12} - \doo_{\nu}\omega_{\mu}^{12} = 2m(\doo_{\mu}v_{s\nu}-\doo_{\nu}v_{s\mu})$}. Although seemingly pure gauge, the latter is in general non-zero by the Mermin-Ho relations Eq. \eqref{eq:Mermin-Ho}. This correspondence is the first major result of the paper. See the SM for a glossary of Riemann-Cartan spacetimes \cite{Shapiro02, Hammond}.

\emph{Conservation laws on curved spacetime.--- } Now we formulate the SF conservation laws in terms of the Weyl fermions coupled to emergent spacetime \cite{NiehYan82b, BradlynRead15, GolanStern18}. The current {$(\rho, J^i)$} in terms of the {Weyl quasiparticles} is $(\tilde{\chi}^\dagger \tau^3\tilde{\chi}, p_F \unitvec{l}^i\tilde{\chi}^\dagger\tilde{\chi} - \frac{1}{2} (\tilde{\chi}^\dagger \im \doo_i\tilde{\chi} -\im\doo_i\tilde{\chi}^\dagger\tilde{\chi}) )$.
To first order in {gradients}, $\doo_{\mu}J^{\mu}$ is equal to $\frac{1}{2} (\tilde{\chi}^\dagger (e^i_1\tau^1 \im\doo_i + e^i_2\tau^2\im\doo_{i})\tilde{\chi} +\hc  )$. On the other hand, this {becomes} (assuming only $\omega_{\mu}^{12} \neq 0$ and $e=$ const.)
\begin{align}
e\doo_{\mu} S^{\mu}_{12}  &= - eT_{12}+eT_{21} \label{eq:Lorentz12}
\end{align}
in terms of the Weyl fermions \cite{GolanStern18}. Here the currents
\begin{align}
T^a_{\ \mu} &= \frac{1}{e} \frac{\delta S_W}{\delta e^{\mu}_a} = \frac{1}{2} \chi^\dagger \tau^a\im \overset{\rightarrow}{D}_{\mu} \chi - \frac{1}{2} \im \chi^\dagger \overset{\leftarrow}{D}_{\mu} \overline{\tau}^a \chi \\
&= \frac{1}{2} (\chi^\dagger \tau^a \im\doo_{\mu}\chi - \im\doo_{\mu}\chi^\dagger \tau^a\chi ) +\frac{1}{8} \omega_{\mu}^{bc} \chi^\dagger \{\tau^a, \sigma_{bc}\}\chi, \nonumber \\
S^{\mu}_{ab} &= 2 \frac{1}{e} \frac{\delta S_W}{\delta \omega_{\mu}^{ab}} = \frac{1}{4} e_c^{\mu}\chi^{\dagger} \{\tau^c ,\sigma^{ab}\} \chi,  
\end{align}
are derived from the relativistic Weyl action Eq. \eqref{eq:Weyl_action}. In particular $S^{\mu}_{12} = \frac{1}{4} \chi^{\dagger}e^{\mu}_a\{\tau^a,\tau^3\} \chi$ and $T^a_{\ b} = e_b^{\mu}T^a_{\mu}$. The relativistic conservation law Eq. \eqref{eq:Lorentz12} follows from the local Lorentz symmetries 
\begin{align}
\delta \chi = -\frac{\im}{4} \Lambda_{ab}\sigma^{ab}\chi 
, \quad \delta e^a_{\mu} = \Lambda^{a}_{\ b} e^b_{\mu}, \nonumber\\
\delta \omega_{\mu}^{ab} = \Lambda^a_{\ c}\omega_{\mu}^{cb} + \omega_{\mu}^{ac} \Lambda_{\ c}^b -\doo_{\mu}\Lambda^{ab},
\end{align} 
which when restricted to the $12$-plane coincide with the Galilean transformations. Indeed the {energy momentum tensor $\Pi^{\mu}_{\nu}$} was not symmetric either in this plane, {the linearization of which is equal to, see the SM},
\begin{align}
\Pi^{\nu}_{\mu} = (\Pi^{(1)} + \Pi^{(2)})^{\nu}_{\mu}=p_F\unitvec{l}^i e j^\nu\delta_{i\mu} - ee^{\nu}_a T^a_{\mu} +e \omega_\mu^{12} S^{\nu}_{12} . \label{eq:lin_energy_momentum}
\end{align}
The Galilean term {$\Pi^{(1)}$} proportional to {$e j^{\nu} = \tilde{\chi}^\dagger e_a^{\nu}\tau^a \tilde{\chi}$} arises due to the finite momentum density {$+p_F\unitvec{l}$} at the node and therefore contributes to {energy-momentum}. {The corresponding relativistic conservation law related to $\Pi^{\mu}_{\nu}$ of the linearized Weyl action follows from} spacetime diffeomorphisms and leads to
\begin{align}
\doo_{\mu} (-e_a^{\mu} T^a_{\nu} + \frac{1}{2}\omega_{\nu}^{ab}S^{\mu}_{ab}) = \doo_{\nu}e^{\mu}_a T^a_{\mu} +\frac{1}{2}\doo_{\nu}\omega_{\mu}^{ab} S^{\mu}_{ab},\label{eq:non_covariant_einstein} 
\end{align}
{i.e. $\doo_{\mu}\Pi^{(2)\mu}_{\nu}= \doo_{\nu}\mathcal{L}$}. The field theory conservation equation for the energy-momentum Eq. \eqref{eq:momentum_conservation} is then equivalent to Eq. \eqref{eq:non_covariant_einstein} and the conservation of the quasiparticle current density $e j^{\mu} = (\tilde{\chi}^\dagger \tilde{\chi},\tilde{\chi}^\dagger (-\mathbf{v}_s^i+v_F\unitvec{l}^i\tau^3)\tilde{\chi})$ at the node (up to subleading terms $O(\doo/p_F)$, {see e.g. \cite{CombescotDombre83, CombescotDombre86}}). 
Althought {the Weyl action \eqref{eq:Weyl_action}} implies the classical conservation law $\doo_{\mu} j^{\mu}= 0$, this suffers from the chiral anomaly at the quantum level due to the emergent spacetime \eqref{eq:tetrad}. 

\emph{Nieh-Yan anomaly. ---}
Adding both chiralities $\pm p_F\unitvec{l}$, the conservation law for momentum is broken since in spite of Eq. \eqref{eq:non_covariant_einstein}, $\doo_{\mu}j_5^{\mu} = \doo_{\mu}(j^{\mu}_+ - j^{\mu}_-) \neq 0$ at the quantum level, i.e. the conservation law suffers from the axial anomaly (however, the {Weyl} qp number is conserved $\doo_{\mu}\sum_{\pm} j_{\pm}^{\mu} = 0$) and leads to the observed momentum non-conservation Eq. \eqref{eq:momentum_anomaly} in the system. The gravitational NY anomaly is \cite{NiehYan82, Yajima96, ChandiaZanelli97, ParrikarEtAl14}, for a chiral pair of Weyl fermions, with $e^a = e^a_{\mu} dx^{\mu}$,
\begin{align}
\doo_{\mu}(e j^{\mu}_5 d^4 x) = \frac{\Lambda^2}{4\pi^2} (T^a \wedge T_a - e^a \wedge e^b \wedge R_{ab}), \label{eq:NYanomaly}
\end{align}
where the  higher order term $O(R^2)$ is neglected. For the $p+ip$ SF, the anomalous chiral Weyl action is \eqref{eq:Weyl_action} with spacetime defined by Eq. \eqref{eq:tetrad}. Note that although the quasiparticles $\tilde{\chi}$ are Majorana-Weyl contributing one-half of Eq. \eqref{eq:NYanomaly} per node, a factor of two comes from accounting for spin degeneracy. {The temporal torsion $T^0 = 0$} and we compute the spatial contribution, 
\begin{align}
T^1 \wedge T^1 + & T^2 \wedge T^2\\
&= 2(\tfrac{c_\parallel}{c_\perp})^2 \epsilon^{0ijk} \left[(\unitvec{l}\cdot \mathbf{v}_s)\doo_i \unitvec{l}_a - \doo_i \mathbf{v}_{sa} \right] \unitvec{l}_j \doo_k \unitvec{l}_a ~d^4 x  \nonumber \\
&=  2(\tfrac{c_\parallel}{c_\perp})^2  \epsilon^{0ijk} \unitvec{l}_i \doo_j \mathbf{v}_s\cdot  \doo_k \unitvec{l}~d^4 x
\approx 0 + O(\doo^3) \nonumber
\end{align}
where $\epsilon^{0xyz}= -1$ and
\begin{align}
T^3 \wedge T^3 &=  \left[2 \epsilon^{0ijk} (\doo_t \unitvec{l}_i-\doo_i(\mathbf{v}_s\cdot \unitvec{l})) \doo_j \unitvec{l}_k \right] d^4 x \nonumber\\
= 2 \epsilon^{0ijk}& \left[ (\doo_t \unitvec{l}_i-(\mathbf{v}_{s}\cdot\nabla)\unitvec{l}_i) \doo_j \unitvec{l}_k - \unitvec{l}_a \doo_i \mathbf{v}_{sa} \doo_j \unitvec{l}_k \right] ~d^4 x  \nonumber\\ 
&\approx 2\epsilon^{0ijk}\left[ (\doo_t \unitvec{l}_i-(\mathbf{v}_{s}\cdot\nabla)\unitvec{l}_i) \doo_j \unitvec{l}_k \right] ~d^4 x + O(\doo^3) \nonumber.
\end{align}
The curvature term $-e_a \wedge e_b \wedge R^{a b}$ is
\begin{align}
&-\tfrac{4\pi}{\kappa}(\tfrac{c_{\parallel}}{c_{\perp}})^{2}\epsilon^{0ijk}  \bigg[  \unitvec{m}_i \unitvec{n}_j (\doo_0 v_{s k}  -\doo_{k}v_{s0}) \nonumber
\\
&+[(\unitvec{m}\cdot\mathbf{v}_s)\unitvec{n}_i - (\unitvec{n}\cdot \mathbf{v}_s)\unitvec{m}_i](\doo_j \mathbf{v}_{sk}) \bigg ] ~d^4x \nonumber\\
&=(\tfrac{c_{\parallel}}{c_{\perp}})^{2}\epsilon^{0ijk} \bigg[2 \unitvec{m}_i \unitvec{n}_j (\unitvec{l} \cdot \doo_t \unitvec{l} \times \doo_k \unitvec{l})\\
&+(\unitvec{l}\times \mathbf{v}_s)_i (\unitvec{l}\cdot \doo_j\unitvec{l}\times \doo_k\unitvec{l}) \bigg] d^4 x . \nonumber
\end{align} 
To lowest order in gradients, we arrive to
\begin{align}
e\doo_{\mu}j^{\mu}_5&= \frac{\Lambda^2}{2 \pi^2}e \left(1-\frac{c_{\perp}^2}{c_{\parallel}^2} \right)\epsilon^{0ijk} [\doo_t \unitvec{l}_i -(\mathbf{v}_s\cdot\nabla)\unitvec{l}_i] \doo_j \unitvec{l}_k  
\end{align}
where $e = (\frac{c_{\parallel}}{c_{\perp}})^{2}$. Matching the expression with the hydrodynamic anomaly \cite{VolovikMineev81, Volovik84, Volovik92} {in the SM, the anisotropic} cut-off is $\Lambda = (\frac{c_{\perp}}{c_{\parallel}})p_F$ and applies in a Weyl SF with the nodes at $\pm p_F\unitvec{l}$, such as $^3$He-A, or in a Weyl SC after minimal substitution \cite{Volovik84, GolanStern18}. The expression is Galilean invariant and the coefficient is proportional to the weak-coupling normal state density ({without the logarithm $\ln (E_F/\Delta_0)$ due to the neglected gapped fermions \cite{VolovikMineev81}}). This the central result of the paper. The NY anomaly equation can be also derived with simple arguments using Landau levels and spectral flow in the case of a torsional magnetic field $T^3_{\mu\nu}$ \cite{Volovik85, BalatskiiEtAl86, CombescotDombre86, ParrikarEtAl14}. {In general}, the dimensional coefficient $\Lambda$ is seen simply to follow from the fact that torsion couples to momentum and that the density of states of the anomalous chiral lowest Landau level branches is momentum dependent. Lorentz invariance would require that the Weyl nodes are {symmetrically} at $p^{\mu}= 0$ which leads to $\Lambda = 0$ at the node. For chiral Weyl nodes with a non-zero separation $2 p^{\mu}$ in momentum space, the coefficient of the torsion anomaly is $\Lambda \propto \abs{p}$ according to the spectral flow calculation.

In condensed matter systems, however, the Weyl description of the quasiparticles and the chiral anomaly breaks down at some cut-off scale. This is in contrast to fundamental Weyl fermions, where the conventional chiral anomalies satisfy IR-UV independence: the anomaly is the same at each energy scale since it can be computed by comparing to a theory with no anomaly simply by adding a high-energy chiral fermion that cancels the anomaly of the original theory \cite{AlvarezGaumeWitten84}.  On the other hand, for $^3$He-A the UV-completion is fully known in terms of the Fermi-liquid theory and the anomalous SF hydrodynamics of $^3$He-A \cite{Volovik2003}. In the idealized $p$-wave BCS pairing model \eqref{eq:action}, the cut-off energy scale $E_{\rm IR}=\Delta$ for the SF is determined from the MF gap equation $c_{\perp} \sim (E_{\rm UV}/p_F)e^{-m_3 n /g} \sim \frac{\Delta}{p_F}$, where $g$ is the $\delta$-function interaction coupling constant and $E_{\rm UV} \sim v_F p_F = c_{\parallel} p_F$ the normal state Fermi energy. The linear quasirelativistic Weyl regime emerges when $E \ll E_{\rm W} = m c_{\perp}^2 = (\frac{c_\perp}{c_\parallel})\Delta$. Therefore the uniaxial anisotropy is simply the relative scale
$\frac{c_{\perp}}{c_{\parallel}} = \frac{E_{\rm IR}}{E_{\rm UV}}$, while the linear Weyl regime is suppressed by an additional factor of $c_{\perp}/c_{\parallel}$ compared to $E_{\rm IR}$ leading to the value of $\Lambda = \frac{c_{\perp}}{c_{\parallel}}p_F$. In $^3$He-A, $c_{\perp}/c_{\parallel}$ is of the order $10^{-3}$ \cite{VollhardtWoelfle}. Remarkably, the hydrodynamic anomaly \eqref{eq:momentum_anomaly}  is the same as in Eq. \eqref{eq:NYanomaly} when all states beyond the linear quasirelativistic Weyl approximation are taken into account. 

\emph{Outlook. --- } We have revisited the anomalous momentum transport in chiral $p$-wave SFs and SCs in terms of a consistent hydrodynamic and low-energy effective theory description. Using the gauge and Galilean symmetries of the system, we have shown how the quasirelativistic Weyl approximation, emergent spacetime and symmetries appear in the semi-classical derivative expansion. The anomalous transport is a consequence of the axial gravitational NY anomaly due to the chiral Weyl fermions on an emergent Riemann-Cartan spacetime with torsion. 

{{In particular}, the emergent spacetime formulation satisfies all the symmetries and conservation laws of the effective field theory required for the gravitational NY anomaly} {and saturates {the} non-zero value from SF hydrodynamics \cite{BevanEtAl97}}. {The early papers \cite{Volovik85, Volovik1986a, BalatskiiEtAl86, Volovik87} treat the anomaly in terms of a momentum space axial gauge field; this follows from our formalism via the subsitution of the tetrad $e_3^{i} = \unitvec{z} + \delta e_3^{i}$, formally equivalent to gauge field $\sim p_F\delta\unitvec{l}_{\mu}$ in the Hamiltonian \cite{ParrikarEtAl14, FerreirosEtAl19, VolovikComment}. In contrast Refs. \cite{KobayashiEtAl18, IshiharaEtAl19} inconsistently consider both contributions independently. Moreover, the emergent gauge field does not correspond to physical symmetry in the system and $p_F$ is an explicit UV scale. This is distinct from the emergent spacetime \eqref{eq:tetrad}, which in addition is valid for arbitrary (semi-classical) textures and superflow, or considerations of other Weyl systems, where the emergent tetrads/gauge fields (e.g. elastic deformations, Fermi velocities, node separation) are independent \cite{GrushinEtAl16, PikulinEtAl16, FerreirosEtAl19}. On the other hand, Ref. \cite{FerreirosEtAl19} sets the NY cut-off to the lattice scale and neglects the breakdown of the linear Weyl spectrum.}

Interestingly, for the emergent spacetime in Eq. \eqref{eq:tetrad}, the anomaly coefficient seems to vanish in the relativistic case $c_{\perp} = c_{\parallel}$ but this is probably an artefact of the break-down of the (weak-coupling) BCS model. Our findings corroborate the subtle interplay of broken Lorentz invariance, anisotropic dispersion, renormalization and the NY-anomaly coefficient $\Lambda$ and should be verified by detailed field theory computations {\cite{ChandiaZanelli97, ParrikarEtAl14, GolanStern18, FlachiEtAl18, CopettiLandsteiner19, ImakiYamamoto19}}. Similarly the relation of the emergent quasirelativistic ({or uniaxial}) spacetime to Newton-Cartan geometries should be clarified \cite{DuvalEtAl85, SonWingate06, Son13, ChristensenEtAl14, Copetti20}. We did not consider the dynamics of the SF order parameter or Goldstone modes, orbital non-analycity \cite{Volovik92, Volovik2003, NissinenVolovik2018} nor derive the Wess-Zumino consistency equation and action for the chiral NY anomaly. This will be a gravitational Chern-Simons term for the tetrad and spin-connection \cite{Volovik1986c, Balatsky87, Dziarmaga02, Zaheed2012}. {Likewise, we did not consider singular vortices, which will lead to additional curvature and zero modes, as well as the Iordanskii force and gravitational Aharonov{-Bohm} phase \cite{Volovik2003} for the quasiparticles}. The connection of emergent spacetime and thermal transport should be explored \cite{GromovAbanov15, KobayashiEtAl18}. {In particular, it is possible that the UV scale $\Lambda$ is supplemented by the IR temperature scale in the anomaly, which can be universal \cite{NissinenVolovik2019}}. These, and other considerations extending previous results in the literature, will be left for the future.

\emph{Acknowledgements. ---}   I wish to thank {S. Fujimoto, and S. Moroz and C. Copetti, partly during the NORDITA
program ``Effective Theories of Quantum Phases of Matter", for useful discussions}. I want to especially thank G.E. Volovik for invaluable feedback and constructive criticism during the course of this work. This work has been supported by the European Research Council (ERC) under the European Union's Horizon 2020 research and innovation programme (Grant Agreement no. 694248).

\clearpage
\onecolumngrid
\begin{center}
\textbf{\large Supplemental Materials: Emergent spacetime and gravitational Nieh-Yan anomaly in chiral $p+ip$ Weyl superfluids and superconductors}
\end{center}
\setcounter{equation}{0}
\makeatletter
\renewcommand{\theequation}{S\arabic{equation}}

{These Supplementary Matearials are organized as follows. In Sec. I we outline the weak coupling BCS model for the chiral p-wave superfluid, its symmetries and conservation laws, followed by the derivation of the hydrodynamic momentum anomaly in Sec. II\ref{sec:Anomaly}. Sec. III\ref{sec:Riemann-Cartan} provides a glossary for Riemann-Cartan spacetimes with torsion. We also outline our conventions regarding relativistic fermions in Sec. IV\ref{sec:Fermions}.} 

Notations {follow the main text}: The order parameter triad is denoted as $\unitvec{m}, \unitvec{n}, \unitvec{l} = \unitvec{m}\times\unitvec{n}$, or as $\hat{e}_1, \hat{e}_2, \hat{e}_3 = \hat{e}_1\times \hat{e}_2$. We treat local relativistic quantities with the Minkowski metric $\eta_{ab} = \textrm{diag}(1,-1,-1,-1)$. The antisymmetric symbol is $\epsilon_{ijk}$ and $\epsilon^{123} = \epsilon_{123} =+1$, whereas $\epsilon_{0123} =+1 = -\epsilon^{0123}$ and the same conventions are extended to the coordinate space. Repeated latin indices {$i,j,k,l$} etc. are summed over with the Euclidean metric.

\section{I.\quad Model, symmetries and conservation laws} \label{sec:Model}
We consider a general $p+\im p$ wave superfluid (SF) on a flat spacetime at zero temperature, focusing on the symmetries and conservation laws . The concrete model we have in mind is chiral $^3$He-A with equal spin pairing \cite{VollhardtWoelfle, Volovik2003, GolanStern18}.
Let $\Psi(x)$ be a spinless fermion $\{\Psi(x), \Psi^\dagger(x')\} = \delta(x-x')$ with equal spin pairing $p$-wave {superfluid}. The action for Grassmann numbers $\{ \Psi(x)$, $\Psi^{\dagger}(x')\} = 0$ is
\begin{align}
S[\Psi,\Psi^{\dagger}] = \int d^4 x~ \Psi^\dagger \im\doo_t \Psi -  \left[ \epsilon(\Psi,\doo_i\Psi) -\frac{g}{4}(\Psi \doo_i \Psi)(\Psi^\dagger \doo_i \Psi^\dagger) \right].
\end{align}
Here the normal state energy is 
\begin{align}
\epsilon(\Psi, \doo_i \Psi) &= \frac{\doo_i\Psi^\dagger \doo_i \Psi}{2m} - \mu \Psi^\dagger \Psi 
\end{align}
where $m$ is the constituent mass, $\mu=p_F^2/2m$ is the chemical potential. In the BCS (mean-field) MF approximation 
\begin{align}
S_{\rm MF}[\Psi,\Psi^{\dagger}, \Delta, \Delta^\dagger] = \int d^4 x~ \Psi^\dagger \im\doo_t \Psi -  \left[ \epsilon(\Psi,\doo_i\Psi) +\frac{1}{2\im}( \Delta^i \Psi^\dagger \doo_i \Psi^\dagger + \Delta^{* i} \Psi \doo_i \Psi ) \right] + \frac{1}{g}\abs{\Delta}^2. \label{seq:action}
\end{align}
The field $\Delta^i(x) = \frac{g}{2\im}\langle \Psi \doo_i \Psi\rangle$ is the MF gap amplitude. The dynamics from the SF free-energy $S_{\rm SF}[\Delta, \Delta^{\dagger}]$ is ignored for the moment and $\Delta(x), \Delta^{\dagger}(x)$ are given background fields.

 From the action Eq. \eqref{seq:action}, the equation of motion for $\Psi(x)$ is
\begin{align}
(\im\doo_t + \mu) \Psi + \frac{\doo^2}{2m}\Psi -\frac{1}{\im} \Delta^i\doo_i\Psi^{\dagger} = 0.
\end{align}
For the chiral $p+ip$-wave SF we have
\begin{align}
\Delta^i = e_1^i -\im e^i_2 = c_{\perp}(\hat{e}_1-i\hat{e}_2)^i = \frac{\Delta_0}{p_F}(\hat{\mathbf{m}}-i\hat{\mathbf{n}})^i, \quad \unitvec{l} = \unitvec{m}\times\unitvec{n}.
\end{align}
The unbroken continuous symmetries of the normal system are U(1)$_N$ $\times$ $\mathbb{R}^3$ $\times$ SO(3)$_{L}$, translation and rotations form a part of the Galilean group. The global U(1)$_N$ symmetry and rotations $R_{\phi_3}$ around $\hat{e}_3 = \hat{e}_1 \times \hat{e}_2$ act as
\begin{align}
\Psi(x) &\to e^{\im \alpha}\Psi(x), \quad \Delta^i(x) \to e^{2\im \alpha}\Delta^i(x) ,\label{seq:U1number}\\
\Psi(x) &\to \Psi(R_{\phi_3}^{-1} x), \quad \Delta^i(x) \to R_{\phi_3} \cdot \Delta^i(R_{\phi_3}^{-1} x) = e^{-\im\phi_3} \Delta^i(R_{\phi_3}^{-1} x),\quad \unitvec{l}(x) \to R_{\phi_3} \unitvec{l}(R_{\phi_3}^{-1} x) \label{eq:3rot}.
\end{align}

\subsection{Conservation laws}
The symmetries imply conservation laws defined from the Noether currents, where under some symmetry transformations with parameter $\lambda$, 
\begin{align}
\Psi(x) \to \Psi(x) + \lambda \delta_\lambda \Psi(x), \quad \delta_{\lambda}\mathcal{L} = \lambda \doo_{\mu}\mathcal{J}_{\lambda}^{\mu},
\end{align}
where the Lagrangian can change by a total derivative $\doo_{\mu}\mathcal{J}^{\mu}_{\lambda}$. The current corresponding to this symmetry is, up to the equations of motion,
\begin{align}
j_{\lambda}^{\mu} = \frac{\doo \mathcal{L}}{\doo(\doo_{\mu} \Psi)} \delta_{\lambda}\Psi - \mathcal{J}^{\mu}_{\lambda} .
\end{align}
We will consider the number current $J^{\mu}$ corresponding to U(1)$_N$ phase transformations Eq. \eqref{seq:U1number}, energy-momentum $t^{\mu}_{\nu}$ corresponding to translations $x^{\mu}\to x^{\mu} + a^{\mu}$ and angular momentum current $S^{\mu 3}$ corresponding to counter clockwise rotations $R_{\phi_3}$ around $e_3$ in Eq. \eqref{eq:3rot}, where $x^{\mu} \to x^{\mu} -\phi_3\epsilon^{t\mu\nu3} x_{\nu}$.  These currents and conservation laws are
\begin{align}
J^{\mu} &= -\im \frac{\doo \mathcal{L}}{\doo (\doo_{\mu}\Psi)} \Psi + \hc
=  (\Psi^{\dagger}\Psi, \frac{\im}{2m}(\doo_i \Psi^{\dagger}\Psi-\Psi^\dagger \doo_i \Psi) ), \\
\doo_{\mu}J^{\mu} &=  -\Delta^i \Psi^{\dagger} \doo_i\Psi^{\dagger} + \Delta^{*i} \Psi\doo_i\Psi,
\end{align}
particle number is not conserved due to the SF order: {the condensate has indefinite number up to Fermion pairs. The expectation value of current is conserved, however, up to the mean-field gap equations.} The energy-momentum current is
\begin{align}
\Pi^{\mu}_{\nu} &= -\frac{\doo \mathcal{L}}{\doo (\doo_{\mu}\Psi)} \doo_{\nu}\Psi + \hc + \mathcal{L} \delta^{\mu}_{\nu},\quad  \doo_{\mu}\Pi^{\mu}_{\nu} = \doo_{\nu}\mathcal{L}\vert_{\rm \Psi, \Psi^\dagger} = \doo_{\nu}\mu \Psi^{\dagger}\Psi - \frac{1}{2\im}(\doo_{\nu} \Delta^i \Psi^\dagger \doo_i\Psi^\dagger + \doo_\nu\Delta^{*i}\Psi \doo_i\Psi ),
\end{align}
where
\begin{align}
\Pi^0_0 &= -\Psi^{\dagger}\im\doo_t \Psi + \mathcal{L} = -\mathcal{H}_{\rm MF},  \quad \Pi^0_i = -\frac{\im}{2} (\Psi^\dagger \doo_i \Psi - \doo_i\Psi^{\dagger}\Psi) = m J^i\\
\Pi^i_{\mu} &= \frac{\doo_i \Psi^{\dagger} \doo_{\mu}\Psi}{2m} +\frac{\doo_{\mu} \Psi^{\dagger} \doo_i\Psi}{2m} + \frac{1}{2\im} \Delta^i\Psi^{\dagger} \doo_{\mu} \Psi^{\dagger} +\frac{1}{2\im} \Delta^{*i}\Psi \doo_{\mu} \Psi, \mu\neq i, \\ 
\Pi^i_i &= \frac{\im}{2} \Psi^\dagger \overset{\leftrightarrow}{\doo_t} \Psi + \mu \Psi^{\dagger}\Psi + \frac{\doo_i \Psi^\dagger \doo_i\Psi}{2m} - \sum_{j\neq i} \frac{\doo_j\Psi^\dagger \doo_j \Psi}{2m} + \frac{1}{2\im}\Delta^j \Psi^\dagger \doo_j \Psi^\dagger +\frac{1}{2\im}\Delta^{*j} \Psi \doo_j \Psi .
\end{align}
Naturally the translation invariance is broken by explicit coordinate dependence in the Lagrangian. Finally, the angular momentum is
\begin{align}
S^{\mu3} &= \Pi^{\mu}_{\nu} \epsilon^{t \nu\lambda3} x_{\lambda}, \quad \doo_{\mu}S^{\mu3} = -\epsilon^{t\nu\lambda}x_{\lambda}\doo_{\nu}\mathcal{L} + \epsilon^{t\nu\lambda3} \Pi^{\nu}_{\lambda}.
\end{align}
For the chiral p-wave SF we have $\Delta^i =\frac{\Delta_0}{p_F}(e_1+ie_2)$, and the angular momentum is not conserved, i.e. $t^{i}_{j}$ is not symmetric, even for constant parameters $\mu, \Delta^i$ since, in the plane perperndicular to $\hat{e}_3$,
\begin{align}
\Pi^{i}_j - \Pi^j_i = +\frac{1}{2\im} (\Delta^i \Psi^\dagger \doo_j \Psi^\dagger - \Delta^{j} \Psi^\dagger \doo_i \Psi^\dagger +\hc ).
\end{align}
There is an unbroken continuous symmetry U(1)$_{N/2+L_3}$, with 
\begin{align}
\doo_{\mu} (\frac{1}{2}J^{\mu} + S^{\mu3}) =& \frac{1}{2}(\Delta^i \Psi^{\dagger} \doo_i\Psi^{\dagger} + \Delta^{*i} \Psi\doo_i\Psi )  + \frac{1}{2\im} (\Delta^1 \Psi^\dagger \doo_2\Psi^\dagger - \Delta^{2} \Psi^\dagger \doo_1 \Psi^\dagger +\hc ) \\
=& \frac{1}{2}\Psi^{\dagger}(\Delta^1 + \im\Delta^2 ) (\doo_1-\im\doo_2)\Psi^{\dagger} + \hc .
\end{align}

\subsection{Linearized quasiparticle {energy-momentum tensor}}
We write $\Psi(x) = \sum_s u_s(x) a_s + v_s^*(x) a_s^\dagger$, $H = \sum_s \varepsilon_s a^\dagger_s a_s$ and linearize the BdG quasiparticles $\Phi$ close to the nodes $\pm p_F\unitvec{l}$ as 
\begin{align}
\Phi(x) = \sum_s \left( \begin{matrix} u_s(x) \\ v_s(x) \end{matrix}\right) a_s  \approx \sum_{\pm} e^{\im p_F\int^x \unitvec{l}(x')\cdot d \mathbf{x}'}\tilde{\chi}_{\pm}(x)
\end{align}
where $\tilde{\chi}_{\pm}(x)$ is slowly varying compared to $p_F$. The energy momentum tensor $\Pi^{\mu}_{\nu}$ in the comoving frame is composed of the terms
\begin{align}
\Pi^0_0 &= -\Psi^{\dagger}\im\doo_t \Psi + \mathcal{L} = -\mathcal{H}_{\rm MF},  \quad \Pi^0_i = -\frac{\im}{2} (\Psi^\dagger \doo_i \Psi - \doo_i\Psi^{\dagger}\Psi) = m J^i\\
\Pi^i_{\mu} &= v^i_s \cdot \frac{\im}{2} (\doo_\mu \Psi^\dagger\Psi-\Psi^\dagger\doo_\mu\Psi) + \frac{\doo_i \Psi^{\dagger} \doo_{\mu}\Psi}{2m} +\frac{\doo_{\mu} \Psi^{\dagger} \doo_i\Psi}{2m} + \frac{1}{2\im} \Delta^i\Psi^{\dagger} \doo_{\mu} \Psi^{\dagger} +\frac{1}{2\im} \Delta^{*i}\Psi \doo_{\mu} \Psi-\mathcal{L}\delta^i_{\mu}.
\end{align}
By the equations of motion $\mathcal{L} \sim -\frac{\doo^2}{4m}(\Psi^\dagger\Psi) \approx 0$. The linearization is found from of the expansions
\begin{align}
\frac{1}{2m}(\doo_{t}\Psi^\dagger \doo_i\Psi + \doo_i\Psi^\dagger \doo_{t}\Psi) &\approx \frac{1}{2} v_F\unitvec{l}^i(\im  \tilde{\chi}^\dagger \tau^3 \doo_t\tilde{\chi} - \im \doo_t \tilde{\chi}^\dagger\tau^3\tilde{\chi}),\\
\frac{1}{2m}(\doo_{j}\Psi^\dagger \doo_i\Psi + \doo_i\Psi^\dagger \doo_{j}\Psi) &\approx  p_F\unitvec{l}^j \tilde{\chi}^\dagger v_F\unitvec{l}^i\tau^3 \tilde{\chi} - \frac{v_F\unitvec{l}^i}{2}(\tilde{\chi}^\dagger \tau^3 \im\doo_j\tilde{\chi}-\im \doo_j\tilde{\chi}^\dagger \tau^3\tilde{\chi} )  \\
\frac{1}{2\im} \Delta^i \Psi^\dagger \doo_\mu \Psi^\dagger + \frac{1}{2\im} \Delta^{*i} \Psi \doo_\mu\Psi &\approx \frac{1}{2}  (\tilde{\chi}^\dagger (e^i_1\tau^1 \im\doo_\mu + e^i_2\tau^2\im\doo_{\mu})\tilde{\chi} +\hc  ) \\ 
\frac{\im}{2} (\doo_\mu \Psi^\dagger\Psi-\Psi^\dagger\doo_\mu\Psi) &= p_F \unitvec{l}^j\tilde{\chi}^\dagger\tilde{\chi}\delta_{\mu j} - \frac{1}{2}
(\tilde{\chi}^\dagger \im \doo_\mu\tilde{\chi} -\im\doo_\mu\tilde{\chi}^\dagger\tilde{\chi}).
\end{align}

\section{II.\quad Chiral superfluid hydrodynamics} \label{sec:Anomaly}
{Here we review the hydrodynamic momentum transport in $p$-wave superfluid/conductor \cite{VolovikMineev81, CombescotDombre86,Volovik92}. Even though a finite normal component exists at $T=0$ due to the gap nodes \cite{VolovikMineev81}, we neglect it for simplicity in the following. In the hydrodynamic expansion, this amounts to $\mathbf{P} = \vek{j}$, mass density $\rho \sim \rho_s$ and $\vek{v}_n =0$ but is sufficient to show the anomaly, Eqs. (1) and (21) in the main text. In addition, we will work to first order in gradients. The problem actually reduces to the famous angular momentum problem of the chiral superfluid/superconductor: even though each Cooper pair carries net angular momentum $\hbar\unitvec{l}$ in a fixed volume, the pairs overlap substantially, where the overlap is of the order of $\Delta_0/p_F v_F$. This makes the local angular momentum density very small and not equal to $\frac{\hbar \rho}{2m}\unitvec{l}$, instead it is of the order of $(\Delta_0/E_F)^2$. At equilibrium, the local variation of angular momentum is still well-defined. Similarly, the linear momentum density is well-defined, although it is no longer conserved \cite{VolovikMineev81, CombescotDombre86}.}  

{The spinless $p$-wave superfluid with order parameter is $\langle \Psi \nabla \Psi \rangle \sim \boldsymbol{\Delta} =  \frac{\Delta_0}{p_F} (\unitvec{m}-\im \unitvec{n})$. The anisotropy direction of Cooper pair orbital angular momentum $\unitvec{l} = \unitvec{m} \times \unitvec{n}$ and $\unitvec{l} \cdot \vek{v}_s$ are the Goldstone variables. We will for now denote the local angular momentum $\vek{L} = L_0 \unitvec{l} = \pm(\frac{\hbar}{2m} \rho - C_0)\unitvec{l}$, where the $\mp$ refers to the convention $\vek{\Delta}\propto (\unitvec{m} \pm \im\unitvec{n})$, and incorporate it as a constraint with Lagrange multiplier $\vek{\Omega}$. 

The energy functional is, $H = \int d^3\vek{x}~ \mathcal{H}$,
\begin{align}
\mathcal{H} = E(\rho, \unitvec{l}, \boldsymbol{v}_s) - \vek{\Omega} \cdot [\vek{L} \mp (\frac{\hbar}{2m}\rho - C_0)\unitvec{l}]
\end{align}
the dynamical constant $C_0 = \vek{L}\cdot \vek{l} \mp \frac{\hbar}{2m}\rho$ has units of angular momentum density and is the semi-classical generator of the combined U(1)$_{L_3\mp N/2}$ gauge symmetry. The hydrodynamic equations follow from, e.g. semi-classical Poisson brackets \cite{Volovik92}. In particular, $\doo_t C_0 = \frac{1}{\hbar}\{H, C_0\} = 0$ and $\doo C_0 \sim 0$. The superfluid London energy $E(\rho, \unitvec{l}, \boldsymbol{v}_s)$ at $T=0$ (for simplicity isotropic in gradients of $\unitvec{l}$ and neglecting spin degree of freedom) is given as
\begin{align}
E(\rho, \vek{v}_s, \unitvec{l}) = \frac{1}{2}\rho \vek{v}_s^2 + \frac{1}{2} K (\nabla_i \unitvec{l}) \cdot (\nabla_i \unitvec{l}) + \vek{v}_s\cdot (\nabla \times C\unitvec{l}) -C_0(\unitvec{l}\cdot \vek{v}_s)(\unitvec{l} \cdot \nabla \times \unitvec{l})
\end{align}
where the coefficients $C = \frac{\hbar}{4m}\rho$ and the dynamical constant $C_0 \simeq \frac{\hbar}{2m}\rho(1 - O(\Delta_0^2/p_F^2 v_F^2))$, the normal state density in weak-coupling theory \cite{Volovik92}.} {The hydrodynamic conservation laws to linear order are \cite{Volovik92},
\begin{align}
\doo_t \rho &= - \nabla \cdot \vek{j} \nonumber\\
\doo_t \vek{v}_s &= -\nabla \mu + \frac{1}{2m}\epsilon_{abc} \unitvec{l}_a \doo_t\unitvec{l}_b \nabla \unitvec{l}_c\\
(\rho - \frac{2m}{\hbar}C_0)\doo_t\unitvec{l} &= \frac{2m}{\hbar}(\frac{\delta E}{\delta \unitvec{l}} \times \unitvec{l}) - (\vek{j}\cdot \nabla)\unitvec{l}, \nonumber
\end{align}
where the mass current \cite{MerminMuzikar80, VollhardtWoelfle}
\begin{align}
\mathbf{P} = \vek{j} = \frac{\delta H}{\delta \vek{v}_s} = \rho \vek{v}_s + \nabla \times (C\unitvec{l}) - C_0\unitvec{l}(\unitvec{l}\cdot\nabla\times  \unitvec{l}).
\end{align} 
The variation of the energy is
\begin{align}
dE = \mu d \rho + T ds + \vek{j} \cdot d \vek{v}_s + \frac{\doo E}{\doo \vek{l}}d \vek{l} + \frac{\doo E}{\doo (\nabla_i \vek{l})} d \nabla_i \vek{l}
\end{align}
The pressure is given as $P = -E + Ts + \mu \rho$. This gives
\begin{align}
dP = -\frac{\doo E}{\doo \boldsymbol{l}}d \boldsymbol{l} -\frac{\doo E}{\doo (\nabla_i \boldsymbol{l})} d \nabla_i \boldsymbol{l} - \vek{j} \cdot d\vek{v}_s + \rho d\mu + s dT.
\end{align}
The superfluid velocity $\vek{v}_s$ and $\unitvec{l}$ are not independent hydrodynamic variables due to the combined gauge symmetry. To see the interdependence of $\unitvec{l}$ and $\vek{v}_s = \frac{\hbar}{2m}\nabla \phi$, the phase rotation $\vek{\Delta} \to e^{\im\phi}\vek{\Delta} = e^{\mp\im \varphi}\vek{\Delta}$ where $\varphi$ is the angle of rotation around $\unitvec{l}$. This gives the variation of $\vek{v}_s$ in terms of $\unitvec{l}$ as $d\vek{v}_s = \frac{\hbar}{4m} (\nabla d\phi - (\unitvec{l}\times \nabla \unitvec{l})\cdot d\unitvec{l})$ and the additional Mermin-Ho relation,
\begin{align}
\nabla \times \vek{v}_s = \pm\frac{\hbar}{4m} \epsilon^{abc} \unitvec{l}_a \nabla \unitvec{l}_b\times \nabla \unitvec{l}_c.
\end{align}
Let us calculate the time derivative of the current, with $\vek{L} = (\frac{\hbar\rho}{2m} - C_0)\unitvec{l}$,
\begin{align}
\doo_t \vek{j} &= \doo_t (\rho \vek{v}_s) + \frac{1}{2} \nabla \times \doo_t (\frac{\hbar\rho}{2m}\unitvec{l}) - C_0 \doo_t(\unitvec{l} ( \unitvec{l} \cdot \nabla\times\unitvec{l})) = -(\nabla \cdot \vek{j})\vek{v}_s - \rho \nabla\mu + \frac{\delta E}{\delta \unitvec{l}_m} \nabla \unitvec{l}_m - (\vek{j}\cdot \nabla)\vek{v}_s + \vek{j}_m\nabla \vek{v}_{s m} \nonumber\\
&+ C_0 \unitvec{l}\cdot \doo_t \unitvec{l} \times \nabla \unitvec{l}  +  \frac{1}{2} \nabla \times \doo_t (\frac{\hbar\rho}{2m}\unitvec{l}) - C_0 \doo_t(\unitvec{l} ( \unitvec{l} \cdot \nabla\times\unitvec{l}))
\end{align}
where we have used the equations of motion and Mermin-Ho relations, with also
\begin{align}
(\vek{j}\cdot \nabla)\vek{v}_s - \vek{j}_m\nabla\vek{v}_{sm} = \frac{\hbar}{2m}\epsilon^{abc} \unitvec{l}_a (\vek{j}\cdot \nabla)\unitvec{l}_b\nabla\unitvec{l}_c .
\end{align}
In addition,
\begin{align}
\doo_t (\unitvec{l} (\unitvec{l}\cdot \nabla \times \unitvec{l})) = 3 \unitvec{l} (\doo_t \unitvec{l}\cdot \nabla \times \unitvec{l}) + \unitvec{l} \cdot \doo_t\unitvec{l} \times \nabla \unitvec{l} - \nabla_j(\unitvec{l} (\unitvec{l} \times \doo_t\unitvec{l})_j) 
\end{align}
The energy-momentum tensor is (see also e.g. \cite{CombescotDombre86})
\begin{align}
\Pi^{ij} = (P-  \nabla_k \unitvec{l}\cdot\frac{\doo E}{\doo (\nabla_k \unitvec{l})})\delta^{ij} + \vek{v}^i_s \vek{j}^j + \nabla_i \unitvec{l}\cdot\frac{\doo E}{\doo (\nabla_j \unitvec{l})} + \frac{1}{2}\epsilon^{ijk} \doo_t ( \frac{\hbar \rho}{2m} \unitvec{l})_k .
\end{align}
As a result, the system of hydrodynamic equations is not closed, and the momentum conservation law is is anomalous
\begin{align}
\doo_t \vek{j}^i + \nabla_j \Pi^{ij} =-3C_0 \unitvec{l}^i \doo_t \unitvec{l} \cdot \nabla \times \unitvec{l}.
\end{align}
For non-zero normal component velocity $\vek{v}_n$, one can use Galilean invariance of $\vek{j}-\vek{j}_s$ \cite{CombescotDombre86}; this amounts to the substitution $\Pi^{ij} \to \Pi^{ij} + B^{ij}$, where
\begin{align}
\nabla_{j}B^{ij} = -3 C_0 (\unitvec{l}_i (\vek{v}_n \cdot \nabla ) \unitvec{l} \cdot \nabla \times \unitvec{l}) + O(\nabla \vek{v}_n) .
\end{align}
The conservation law becomes
\begin{align}
\doo_t \vek{j}^i + \nabla_j \Pi^{ij} =-3C_0 \unitvec{l}^i (\doo_t - \vek{v}_s\cdot\nabla) \unitvec{l} \cdot \nabla \times \unitvec{l} = -\doo_t \vek{P}_{\rm qp} - \nabla_j \Pi_{\rm qp}^{ij}
\end{align}
up to $O(\doo v_{s})$ terms with the identification $\vek{v}_{n} = -\vek{v}_s$ in the comoving frame. We emphasize that $C_0 \simeq \frac{\hbar\rho}{2m} = \frac{1}{2}\frac{k_F^3}{3\pi^2, \hbar^2}$ with \cite{VolovikMineev81}
\begin{align}
\frac{\hbar \rho}{2m} - C_0 = \frac{1}{8m}\rho(\tfrac{\Delta_0}{E_F})^2 \ln (\tfrac{E_F}{\Delta_0})
\end{align}
at $T=0$ in the absence of Fermi-liquid corrections. $C_0$ is a parameter fixed by the normal state density; for the Bose systems of small molecules, $C_0 =0$ and the anomaly vanishes \cite{VolovikMineev81,Volovik92}. The anomaly expression is valid for arbitrary textures and superflow, to the first order in the small gradients. On the other hand, for the normal component excitations, the hydrodynamic regime is not valid due to the much longer mean-free path at any temperature $\ell$ \cite{CombescotDombre86}. Specifically, $\ell/L \sim L/\xi \gg 1$, where $L$ is the scale of hydrodynamic variations and $\xi = \hbar v_F/\Delta$ the coherence length.} 

\section{III.\quad Riemann-Cartan spacetime} \label{sec:Riemann-Cartan}
Riemann-Cartan spacetimes are spacetimes with non-zero torsion, i.e. manifolds equipped with antisymmetric metric compatible affine connection in the tangent space. Here we provide a glossary of the relevant terms utilized in the main text, see e.g. \cite{NiehYan82b, ParrikarEtAl14, Shapiro02, Hammond}. The spacetime is defined by a four manifold $M$, tetrad field, spin-connection and a metric compatible connection. Equivalently we can speak of the torsion and curvature on the manifold.

They both enter in the commutator of covariant derivatives
\begin{align}
[\nabla_{\mu},\nabla_{\nu}] V^{\lambda} = -T^{\rho}_{\mu \nu} \nabla_{\rho}V^{\lambda} + R^{\lambda}_{\mu \nu \rho} V^{\rho}.
\end{align}
The geometric interpretation of torsion is the non-closure of infinitesimal paralleogram lifted to the tangent space, whereas curvature corresponds to the rotation of tangent vectors along loops in the tangent space.

Alternatively $e^a_{~\mu}(x)$ is the vielbein or tetrad field 
\begin{align}
g_{\mu\nu}(x) = \eta_{ab} e^{a}_{~\mu}(x) e^b_{~\nu}(x), \quad e^{~\mu}_{b} e^a_{~\mu} = \delta^{a}_{b}, 
\end{align}
that defines an orthonormal Lorentz frame $e_a \equiv e_a^{~\mu}\doo_{\mu}$ with indices $a=0,1,2,3$ in the tangent space. The one-form $e^a = e^a_{\mu} dx^{\mu}$ is the dual. The Lorentz and coordinate indices are lowered with $\eta_{ab}$ and $g_{\mu\nu}$ respectively, and basis transformations with $e^a_{~\mu}$ and $e^{~\mu}_a$. The connection is defined by the different connection coefficients when taking derivatives of mixed tensors 
\begin{align}
\nabla_{\mu} X^a_{~\nu} &\equiv \doo_{\mu}X^a_{\nu} + \omega^a_{\mu b} X^b_{~\nu} - \Gamma^{\lambda}_{\mu\nu}X^a_{~\lambda} \\
\nabla_{\mu} X_a^{~\nu} &\equiv \doo_{\mu}X_a^{~\nu} - \omega^b_{\mu a} X_b^{~\nu} + \Gamma^{\nu}_{\mu\lambda}X_a^{~\lambda}.
\end{align}
where $a,b$ are Lorentz indices and  $\Gamma^\lambda_{\mu\nu}$ are the connection coefficients in the tangent space coordinate basis $\doo / \doo x^{\mu}$. Note that both the tetrad $e^{a}_{~\mu}$ (i.e. the metric) and the spin-connection $\omega^{ab}_{\mu}$ (or the connection $\Gamma^\lambda_{\mu\nu}$) are independent geometric variables that enter e.g. the Dirac equation for fermions.  The connection $\Gamma^{\lambda}_{\mu\nu}$  is not independent of $e^a_{~\mu}$ and $\omega^{ab}_{\mu}$, since the tetrad is defined to be covariantly constant
\begin{align}
\nabla_{\mu}e^a_{~\nu} = 0, \quad \nabla_{\mu} e^{~\lambda}_{a} =0.
\end{align}
The spin-connection is solved as
\begin{align}
\omega_{\mu b}^{a} = e_b^{~\nu}(\Gamma^{\lambda}_{\mu\nu}e^a_{~\lambda}-\doo_{\mu}e^a_{~\nu}).
\label{eq:spin_connection}
\end{align}
In addition to Eq. \eqref{eq:spin_connection} it follows that $\omega_{\mu ba} = -\omega_{\mu ab}$ and  $\nabla_{\lambda} g_{\mu\nu} =0$ i.e. the connection is also metric compatible. Further, without loss of generality, we have
\begin{align}
\Gamma^{\lambda}_{\mu\nu} &= \mathring{\Gamma}^{\lambda}_{\mu\nu} + C^{\lambda}_{\mu\nu},  \quad \mathring{\omega} = \omega(\mathring{\Gamma},e)\\
\mathring{\Gamma}^{\lambda}_{\mu\nu} &= \tfrac{1}{2}g^{\lambda\rho}(\doo_{\mu}g_{\rho \nu} + \doo_{\nu}g_{\mu\rho}-\doo_{\rho}g_{\mu\nu}), \quad
C_{\lambda\mu\nu} = \tfrac{1}{2}(T_{\lambda\mu\nu} - T_{\mu\nu\lambda} + T_{\nu\lambda\mu})
\end{align}
where $C_{\lambda\mu\nu} \equiv g_{\lambda\rho} C^{\rho}_{\mu\nu}$ is the contorsion tensor in terms of contractions of the torsion tensor $T_{\lambda\mu\nu}\equiv g_{\lambda\rho}T^{\lambda}_{\mu\nu}$. 

The tetrad defines a one-form field $e^a = e^a_{\mu}dx^{\mu}$ and the torsion and curvature follow in differential form notation as
\begin{align}
T^a &= d e^a + \omega^a_{\ b} \wedge e^b = \frac{1}{2} T^a_{\mu\nu}  dx^{\mu} \wedge dx^{\nu} \\
R^a_{ \ b} &= d \omega^a_{\ b} + \omega^a_{\ c} \wedge \omega^{c}_{\ b} = \frac{1}{2} R^a_{\mu\nu b} dx^{\mu} \wedge dx^{\nu}. 
\end{align}
They satisfy the Bianchi identities
\begin{align}
d T^a + \omega^a_{~b} \wedge T^b = R^a_{~b} \wedge e^b, \quad d R^a_{~b} + \omega^a_{~c} \wedge R^c_{~b} + R^a_{~c} \wedge \omega^c_{~b} = 0.  
\end{align}
These lead to the Nieh-Yan form as $d(e^a \wedge T_a) = T^a \wedge T_a - e^a \wedge e^b \wedge R_{ab}$ \cite{NiehYan82}.  Their coordinate expression are
\begin{align}
T^{\lambda}_{\mu \nu} = \Gamma^{\lambda}_{\mu \nu} - \Gamma^{\lambda}_{\nu \mu}, \quad 
R^{\lambda}_{\mu\nu \rho} = \doo_{\mu} \Gamma^{\lambda}_{\nu \rho} + \Gamma^{\lambda}_{\mu\sigma}\Gamma^{\sigma}_{\nu\rho} - (\mu \leftrightarrow \nu)
\end{align}
where $R^{\lambda}_{\mu\nu \rho}$ is the Riemann curvature tensor. Note that $\mathring{R}^{ab} \neq R^{ab}$ due to torsion. We have the identities
\begin{align}
e^{-1} \doo_{\mu} e = \mathring{\Gamma}^{\lambda}_{\mu \lambda} = \Gamma_{\mu\lambda}^{\lambda} - C^{\lambda}_{\mu\lambda} = \mathring{\Gamma}_{\lambda\mu}^\lambda= \Gamma^{\lambda}_{\lambda\mu}-T^{\lambda}_{\lambda\mu}
\end{align}
which are useful in the case of the spacetime Eq. \eqref{eq:tetrad} in the main text where $e$ is constant.

\section{IV.\quad Relativistic fermions} \label{sec:Fermions}
We summarize our conventions on relativistic fermions on curved spaces with torsion \cite{NiehYan82b, ParrikarEtAl14, BradlynRead15}. Let $\psi$ be a Dirac fermion.  Under a (local) Lorentz transformation $x \to x'=\Lambda x$, $\psi$ transforms
\begin{align}
\psi(x)  \to \psi'(x') =  S(\Lambda) \psi(x)
\end{align}
where
\begin{align}
\Lambda &= e^{-\im \lambda_{ab}L^{ab}/2}, \quad L^{ab}_{\ \ \mu\nu} = \im(\delta^a_{\ \mu}\delta^b_{\ \nu} - \delta^a_{\ \nu} \delta^{b}_{\ \mu}), \label{seq:LorentzRep} \\
S(\Lambda) &= e^{-\im \lambda_{ab}\sigma^{ab}/4}, \quad \sigma^{ab} = \frac{\im}{2} [\gamma^a,\gamma^b]. 
\end{align}
The gamma matrices in the Weyl basis are  
\begin{align}
\gamma^{a} = \left( \begin{matrix} & \sigma^{a} \\ \overline{\sigma}^{a} \end{matrix} \right), \quad \gamma^5 = \im \gamma^0\gamma^1\gamma^2\gamma^3 = \left( \begin{matrix} -1 &  \\ & 1 \end{matrix} \right), \quad \{\gamma^a, \gamma^b \} =2\eta^{ab}
\end{align}
where $\sigma^{\mu} = (1, \sigma^i)$ and $\overline{\sigma}^\mu = (1,-\sigma^i)$ and $\eta^{ab} = \textrm{diag}(1,-1,-1,-1)$. The massless Dirac action is, with $\overline{\psi}=\psi^\dagger \gamma^0$,
\begin{align}
S_{\rm Dirac} = \frac{1}{2} \int e d^4 x~  \overline{\psi} e_a^{\mu}\gamma^a  \im\overset{\rightarrow}{D}_{\mu}  \psi - \overline{\psi}\im \overset{\leftarrow}{D}_{\mu}e_a^{\mu}\gamma^a  \psi ,
\end{align}
where $\overset{\rightarrow}{D}_{\mu} = \doo_{\mu} - \frac{\im}{4} \omega_{\mu}^{ab}\sigma_{ab}$ and $\overset{\leftarrow}{D}_{\mu} = \doo_{\mu} + \frac{\im}{4} \omega_{\mu}^{ab}\sigma_{ab}$ acts on the hermitean conjugate from the right. We write $\psi = (\psi_{\rm L} \ \psi_R )$ with $\gamma^5 \psi_{\rm R,L} = \pm \psi_{\rm R,L}$. The gamma matrices $\gamma^0 \gamma^a =  (\sigma_{\rm L}^a, \sigma_{\rm R}^a)=(\overline{\sigma}^a, \sigma^a)$ are block diagonal and $\sigma^{0i} = (\im\overline{\sigma}^i,\im\sigma^i)$ and $\sigma^{ij} = \epsilon^{ijk}\sigma^k$. The action becomes
\begin{align}
S_{\rm Weyl} = \sum_{\rm R,L}\frac{1}{2} \int e d^4 x~  \psi^\dagger_{R,L} e_a^{\mu}\sigma_{\rm R,L}^a  \im\overset{\rightarrow}{D}_{\rm R,L \mu}  \psi_{\rm R,L} - \psi^\dagger_{\rm R,L} \im\overset{\leftarrow}{D}_{\rm R,L \mu}e_a^{\mu}\sigma_{\rm R,L} ^a  \psi_{\rm R,L} .
\end{align}
This expression is invariant under local Lorentz transformations for which $\omega_{\mu} =\frac{1}{4} \omega_{\mu}^{ab} \sigma_{ab}$ is a connection, since
\begin{align}
\omega_{\mu} \to S(\Lambda)\omega_{\mu}S(\Lambda)^{-1}  + S(\Lambda)\im\doo_{\mu}S(\Lambda)^{-1} .
\end{align}
Similarly, the transformation law $S(\Lambda)^{\dagger}\tau^aS(\Lambda) = \Lambda^a_{\ b}\tau^b$ follows from the representation in Eq. \eqref{seq:LorentzRep}.
\subsection{Conservation laws on curved spaces}
The conservation laws in the main text are written in a form where the (quasi)relativistic symmetries are not manifest. Now we derive the conservation laws due to local Lorentz and spacetime diffeomorphisms. The local Lorentz transformations
\begin{align}
\delta e^a_{\mu} = \Lambda^{a}_{\ b} e^b_{\mu},\quad \delta \omega_{\mu}^{ab} = \Lambda^a_{\ c}\omega_{\mu}^{cb} + \omega_{\mu}^{ac} \Lambda_{\ c}^b -\doo_{\mu}\Lambda^{ab} ,\quad \delta \chi = -\frac{\im}{4} \Lambda_{ab}\sigma^{ab}\chi 
\end{align} 
which give the equation $D^\omega_{\mu} (eS^{\mu}_{ab}) = e(T_{ab}-T_{ba})$, where $D^{\omega}_{\mu}(eS^{\mu}_{ab}) \equiv \doo_{\mu}(e S^{\mu}_{12})-\omega_{\mu a}^{c} eS^{\mu}_ {cb} -\omega_{\mu b}^{c}eS^{\mu}_{ac}$.  This is the covariant form of \eqref{eq:Lorentz12} in the main text.

For the spacetime transformations $x^{\mu} \to x^{\mu}+\xi^{\mu}$, 
\begin{align}
\delta e^{\mu}_a &= \doo_{\nu}\xi^\mu e^\nu_a - \xi^{\nu}\doo_{\nu}e^{\mu}_a = \nabla_{\nu}\xi^{\mu}e_a^{\nu} - e_a^{\nu}e_b^{\mu} T^b_{\nu\lambda}\xi^{\lambda}+\delta_{\xi\cdot \omega}e_a^{\mu}, \\ 
\delta \omega^{ab}_{\mu} &=  -\doo_{\mu}\xi^\nu \omega_\nu^{ab} - \xi^{\nu}\doo_{\nu}\omega_{\mu}^{ab} = R^{ab}_{\mu\nu}\xi^{\nu}+\delta_{\xi\cdot \omega}\omega_{\mu}^{ab}
\end{align}
where $\delta_{\xi\cdot\omega}$ is a Lorentz transformation by $\delta \Lambda^{ab} = \xi^{\mu}\omega_{\mu}^{ab}$. The conservation law is, up to local Lorentz transformations,
\begin{align}
\nabla_{\mu} (e_a^{\mu} T^a_{\nu}) + T^{\lambda}_{\mu\nu}(e^{\mu}_a T^a_{\lambda}) = +\frac{1}{2}R^{ab}_{\nu\mu}S^\mu_{ab}.
\end{align}
which is the covariant form of Eq. \eqref{eq:non_covariant_einstein} in the main text.

\end{document}